\documentclass[10pt,twoside]{hsqcd}
\usepackage{epsf,amsmath}
\usepackage{graphicx}

\begin{document}

\title{DOUBLE PARTON CORRELATIONS IN PERTURBATIVE QCD \\}

\author{\underline{A.~M.~Snigirev} \\ \\
 D.~V.~Skobeltsyn Institute of Nuclear Physics,\\
M.~V.~Lomonosov  Moscow State University\\
Moscow, 119991, Russia\\
E-mail: snigirev@lav01.sinp.msu.ru }

\maketitle

\begin{abstract}
\noindent This talk brings attention to what is knowable from perturbative
QCD theory on two-parton distribution functions in the light of  CDF measuruments of the inclusive  cross section for double parton scattering.
\end{abstract}



\markboth{\large \sl \underline{A.M. Snigirev} 
\hspace*{2cm} HSQCD 2008} {\large \sl \hspace*{1cm} DOUBLE PARTON CORRELATIONS IN PERTURBATIVE QCD}

The Collider Detector at Fermilab (CDF) Collaboration has  measured a large number of double parton scattering~\cite{cdf}  providing new  and complementary information on the structure of the proton and  parton-parton correlations. 
The possibility of observing two separate hard  collisions has been
proposed since long~\cite{landshoff}, and from that has also 
developed in a number of works~\cite{takagi, trelani}. 
A brief review of
the current situation and some progress in the modeling account of 
correlated flavour, colour,
longitudinal and transverse momentum distributions can be found
in Ref.~\cite{sjostrand2}. Multiple interactions require an ansatz for the structure
of the incoming beams, i.e. correlations between the constituent partons. As a
simple ansatz, usually,  the two-parton distributions are supposed to be 
the product of two single-parton distributions times a momentum conserving 
phase space factor. In recent papers~\cite{snig03} it has been shown that this
hypothesis is in some contradiction with the leading logarithm approximation
of perturbative QCD  (in the framework of
which a parton model, as a matter of fact, was established in the quantum field 
theories~\cite{gribov}). 

In order to be clear and to introduce the denotations
let us recall that,
for instance, the
differential cross section for the four-jet process (due to the simultaneous
interaction of two parton pairs) is given by~\cite{takagi}  
\begin{equation} 
\label{fourjet}
d \sigma = \sum \limits_{q/g} \frac{ d \sigma_{12} ~d \sigma_{34}}
{\sigma_{\rm eff}}~ D_ p(x_1,x_3)~D_{\bar{p}}(x_2,x_4), 
\end{equation} 
where $d \sigma_{ij}$ stands for the two-jet cross section. The dimensional
factor $\sigma_{\rm{eff}}$
in the denominator represents the total inelastic cross section which is an
estimate of the size of the hadron, $\sigma_{\rm eff} ~\simeq~2 \pi r_ p^2$ (the factor
2 is introduced due to the identity of the two parton processes). With
the effective cross section measured by CDF, $(\sigma_{\rm eff})_{\rm CDF}=
(14.5 \pm 1.7^{+1.7}_{-2.3})$ mb~\cite{cdf}, one can estimate the transverse
size $r_p~\simeq 0.5$ fm, which is too small in comparison with the proton
radius $R_p$
extracted from $ep$ elastic scattering experiments. The relatively small
value of $(\sigma_{\rm eff})_{\rm CDF}$ with respect to the naive expectation
$2 \pi R_ p^2$ was, in fact, considered~\cite{trelani} as  evidence of
nontrivial correlation effects in transverse space. But, apart from these
correlations, the longitudinal momentum correlations can also exist and 
they were   investigated in Ref.~\cite{snig03}. The factorization ansatz is just applied
to the two-parton distributions incoming in Eq.~(\ref{fourjet}):
\begin{equation} 
\label{factoriz}
D_ p(x_i,x_j)~ = ~ D_ p(x_i,Q^2)~ D_ p(x_j,Q^2)~(1-x_i-x_j),
\end{equation} 
where  $D_ p(x_i,Q^2)$ are the single quark/gluon momentum distributions at
the scale $Q^2$ (deter\-mi\-ned by a hard process).

However many parton distribution functions satisfy the generalized 
Gribov-Lipatov-Altarelli-Parisi-Dokshitzer (GLAPD) 
evolution equations derived for the first
time in Refs~\cite{kirschner,snig} as well as single
parton distributions satisfy more known and cited GLAPD
equations~\cite{gribov,altarelli}.
Under certain initial conditions these generalized
equations lead to  solutions, which are identical with the jet calculus rules
proposed originally for multiparton fragmentation functions by
Konishi-Ukawa-Veneziano~\cite{konishi} and are in some contradiction with the
factorization hypothesis (\ref{factoriz}). Here one should note that at the parton  
level this  is the strict assertion within the leading logarithm approximation. 

After introducing the natural dimensionless variable
$$t = \frac{1}{2\pi b} \ln \Bigg[1 + \frac{g^2(\mu^2)}{4\pi}b
\ln\Bigg(\frac{Q^2}{\mu^2}\Bigg)\Bigg]~=~\frac{1}{2\pi b}\ln\Bigg
[\frac{\ln(\frac{Q^2}{\Lambda^2_{\rm QCD} })}
{\ln(\frac{\mu^2}{\Lambda^2_{\rm QCD}})}\Bigg]
,~~~~~b = \frac{33-2n_f}{12\pi}~~
{\rm {in~ QCD}},$$
where $g(\mu^2)$ is the running coupling constant at the reference scale
$\mu^2$, $n_f$ is the number of active flavours, 
$\Lambda_{\rm QCD}$ is the dimensional QCD 
parameter,
the GLAPD equations read~\cite{gribov,altarelli}
\begin{equation}
\label{e1singl}
 \frac{dD_i^j(x,t)}{dt} = 
\sum\limits_{j{'}} \int \limits_x^1
\frac{dx{'}}{x{'}}D_i^{j{'}}(x{'},t)P_{j{'}\to j}\Bigg(\frac{x}{x{'}}\Bigg).
\end{equation}
\noindent
They describe the scaling violation 
of the parton distributions $ D^j_i(x,t)$ inside a dressed quark or gluon ($i,
j~=~q/g$).

We will not write  the kernels $P$ explicitly and  derive the generalized 
equations for  two-parton distributions
$D_i^{j_1j_2}(x_1,x_2,t)$, representing the probability that in a dressed
constituent $i$ one finds two bare partons  of  types $j_1$ and $j_2$ with 
the given longitudinal momentum fractions $x_1$ and $x_2$  (referring to
 ~\cite{snig03,gribov,kirschner, snig, altarelli} for details), we 
note only that their solutions can be represented as the convolution of single 
distributions~\cite{kirschner, snig}.
This convolution coincides with the jet calculus rules~\cite{konishi} as
mentioned above and is the  generalization of 
the  well-known Gribov-Lipatov relation
installed for single functions~\cite{gribov} (the distribution
of bare partons inside a dressed constituent  is identical to the distribution 
of dressed constituents in the fragmentation of 
a bare parton in the leading logarithm 
approximation). 
The obtained solution  shows also that the double distribution of partons is 
{\it {correlated}} in the leading logarithm approximation:
\begin{eqnarray}
\label{nonfact}
D_i^{j_1j_2}(x_1,x_2,t) \neq D_{i}^{j_1}(x_1,t) 
D_{i}^{j_2}(x_2,t).
\end{eqnarray}

Of course, it is interesting to find out the phenomenological issue of this
parton level consideration. This can be done within the well-known
factorization of soft and hard stages (physics of short and long
distances). As a result the equations (\ref{e1singl}) 
 describe the evolution of parton distributions in a hadron with
$t ~(Q^2)$, if one replaces the index $i$ by index $h$ only. However, the initial
conditions for new equations at $t=0 ~(Q^2=\mu^2)$ are unknown a priori and must
be introduced phenomenologically or must be extracted from experiments 
or some models dealing with physics of long distances [at the parton level: 
$D_{i}^{j}(x,t=0)~= ~\delta_{ij} \delta(x-1)$; ~$D_i^{j_1j_2}(x_1,x_2,t=0)~=~0$].
Nevertheless the solution of the generalized  
GLAPD evolution equations with the given initial
condition may be written as before via the convolution of single
distributions~\cite{snig03,snig}.
This result shows that
if the two-parton distributions are factorized at some scale $\mu^2$, then
the evolution violates this factorization {\it{ inevitably}} at any different
scale ($Q^2 \neq \mu^2$), apart from the violation due to 
the kinematic correlations induced by the momentum
conservation.

For a practical employment it is interesting to know the degree of this
violation. 
Partly this problem was investigated theoretically in Refs.~\cite{snig, snig2}
and  for the two-particle correlations of fragmentation
functions in Ref.~\cite{puhala}. That technique is based on the Mellin 
transformation of distribution functions and the asymptotic behaviour can be estimated.
Namely, with the growth of $t~(Q^2)$ the correlation term  becomes {\it {dominant}} 
for  finite $x_1$ and $x_2$~\cite{snig2} and
thus the two-parton distribution functions 
``forget'' the initial conditions unknown a
priori and the correlations perturbatively calculated appear.

The asymptotic prediction ``teaches'' us a tendency only and tells nothing about the 
values of $x_1,x_2, t(Q^2)$ beginning from which the correlations are significant. 
Naturally numerical estimations
can give an answer to this specific question. We do it using the CTEQ 
fit~\cite{cteq} for single distributions as an input.
The nonperturbative initial conditions $D_h^j(x,0)$ are specified in a parametrized
form at a fixed low-energy scale $Q_0=\mu=1.3$ GeV. The particular function forms and
the value of $Q_0$ are not crucial for the CTEQ global analysis at the flexible
enough parametrization.
The results of numerical calculations are presented in Fig.~1 for the ratio:
\begin{equation}
\label{ratio}
R(x,t)~=~\Big(D_{p(QCD)}^{gg}(x_1,x_2,t) \Big/ D_p^{g}(x_1,t)D_p^{g}(x_2,t)(1 - x_1
- x_2)^2\Big)\Big|_{x_1=x_2=x}.
\end{equation}

\begin{figure}[!thb]
\vspace*{7.0cm}
\begin{center}
\includegraphics{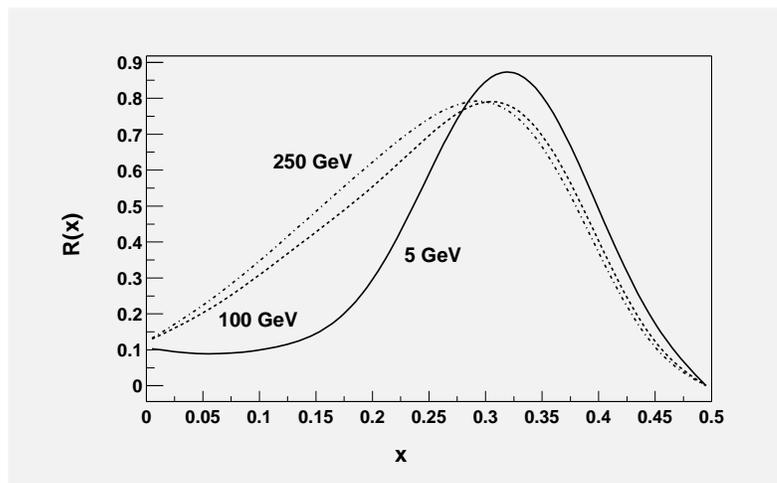}
\caption[*]{ The ratio of perturbative QCD correlations to the factorized component for
the double gluon-gluon distribution in the proton as a function of $x = x_1 = x_2$
for three values of $Q$ = 5 (solid), 100 (dashed), 250 (dash-dotted) GeV.}
\end{center}
\label{fig1}
\end{figure}

Figure~1 shows that
at the scale of CDF hard process ($\sim 5$ GeV)
the ratio (\ref{ratio}) 
 is nearly 10$\%$  and increases right up to 30$\%$ at the LHC 
scale ($\sim 100$ GeV) for the longitudinal momentum fractions $x \leq 0.1$
accessible to these measurements. For the finite longitudinal momentum
fractions $x \sim 0.2 \div 0.4$ the correlations are large right up to 90$\%$ .
They become important in more and more $x$ region   with the growth of $t$
in accordance with the predicted QCD asymptotic behaviour~\cite{snig, snig2}.

The correlation effect is strengthened insignificantly (up to 2$\%$)
for the longitudinal momentum fractions $x \leq 0.1$ when
starting from the slightly lower value $Q_0 = 1$ GeV (early used by CTEQ 
Collaboration). We conclude also that $R(x,t)\rightarrow \rm const$ at $x \rightarrow 0$
most likely, calculating this ratio ($\simeq 0.1$) at $x_{\rm min} = 10^{-4}$.

Seemingly the correction to the double gluon-gluon distributions at the CDF
scale can be smoothly absorbed by uncertainties in the $\sigma_{\rm eff}$ increasing
the transverse effective size $r_p$ by a such way. But this augmentation is still 
not enough to solve a problem of the relatively small value of $r_p$ with 
respect to the proton radius without  
nontrivial correlation effects in transverse space~\cite{trelani}.

Recently 
a nonminor role of the QCD evolution of multiparton distribution
functions  has been also demonstrated~\cite{del05}.
In the case of multiple production of 
$W$ bosons with equal sign, the terms with
correlations may represent a correction of the order of 40$\%$ of the cross
sections,
for $pp$  collisions at 1 TeV c.m. energy,  and a correction of the order of 
20$\%$ at 14 TeV.
In the case of $b{\bar b}$ pairs the correction terms are of the order of
10-15$\%$ at 1 TeV and of the order of 5$\%$ at 14 TeV.


In summary, the numerical estimations show that the leading logarithm perturbative
QCD correlations are quite comparable with the factorized distributions. With
increasing a number of observable multiple collisions (statistic) the more precise
calculations of their cross section (beyond the factorization hypothesis) will be
needed also. In order to obtain the more delicate their characteristics
(distributions over various kinematic variables) it is desirable to implement the
QCD evolution of two-parton distribution functions in some Monte Carlo event
generator as this was done for single distributions. 

\section*{Acknowledgements} 

Discussions with E.E.~Boos, M.N.~Dubinin, V.A.~Ilyin, V.L.~Korotkikh, L.N.~Lipatov, I.P.~Lokhtin, S.V.~Molodtsov, A.S.~Proskuryakov,
L.I.~Sarycheva, V.I.~Savrin, T.~Sjostrand, D.~Treleani 
and G.M.~Zinovjev are gratefully acknowledged. Author is specially thankful to the organizers of HSQCD 2008 for the warm welcome and hospitality.
This work is partly supported 
by Russian Foundation for Basic Research (grants No 08-02-91001 and No 08-02-92496)
and Grants of President of Russian Federation for support of Leading Scientific Schools (No 1456.2008.2).

\end{document}